\documentclass[english,aps,prb,showpacs,superscriptaddress, preprint]{revtex4}
\usepackage[T1]{fontenc}
\usepackage[latin9]{inputenc}
\usepackage{array}
\usepackage{longtable}
\usepackage{amsmath}
\usepackage{graphicx}
\usepackage{amssymb}

\makeatletter

\newcommand{\lyxmathsym}[1]{\ifmmode\begingroup\def\b@ld{bold}
  \text{\ifx\math@version\b@ld\bfseries\fi#1}\endgroup\else#1\fi}

\providecommand{\tabularnewline}{\\}

\@ifundefined{textcolor}{}
{%
 \definecolor{BLACK}{gray}{0}
 \definecolor{WHITE}{gray}{1}
 \definecolor{RED}{rgb}{1,0,0}
 \definecolor{GREEN}{rgb}{0,1,0}
 \definecolor{BLUE}{rgb}{0,0,1}
 \definecolor{CYAN}{cmyk}{1,0,0,0}
 \definecolor{MAGENTA}{cmyk}{0,1,0,0}
 \definecolor{YELLOW}{cmyk}{0,0,1,0}
 }

\makeatother

\usepackage{babel}

\begin{document}

\title{Quantum molecular dynamics simulations of lithium melting using Z-method}
\author{Dafang Li}
\affiliation{LCP, Institute of Applied Physics and Computational
Mathematics, Beijing 100088, People's Republic of China}
\author{Ping Zhang}
\thanks{zhang\_ping@iapcm.ac.cn}
\affiliation{LCP, Institute of Applied Physics and Computational
Mathematics, Beijing 100088, People's Republic of China}
\affiliation{Center for Applied Physics and Technology, Peking
University, Beijing 100871, People's Republic of China}
\author{Jun Yan}
\thanks{yan\_jun@iapcm.ac.cn}
\affiliation{LCP, Institute of Applied Physics and Computational
Mathematics, Beijing 100088, People's Republic of China}
\affiliation{Center for Applied Physics and Technology, Peking
University, Beijing 100871, People's Republic of China}


\begin{abstract}
We performed first-principles molecular dynamics calculations for
lithium using the projector augmented waves method and the
generalized gradient approximation as exchange-correlation energy.
The melting curve of lithium was computed using the
\textit{Z}-method technique for pressures up to 30 GPa, which agrees
well with the experimental and two-phase simulated results. The
change of the melting line slope from positive to negative was
predicted by the characteristic shape inversion of the \textit{Z}
curve at about 8.2 GPa. Through analyzing the static properties, we
conclude that no liquid-liquid phase transition accompanies the
occurrence of the melting line maximum, which is caused by the
higher compressibility of the liquid phase compared to the solid
phase. In addition, we systematically studied the dynamic and
optical properties of lithium near melting curve at critical
superheating and melting temperatures. It was suggested that spectra
difference at critical superheating and melting temperature may be
able to diagnose the homogeneous melting.
\end{abstract}
\maketitle

\section{Introduction}

The simple alkali metals like Li and Na have drawn extensive
attentions recently due to their enigmatic melting behavior at high
pressure. Intensive theoretical and experimental studies have
demonstrated that they undergo a sequence of symmetry-breaking
structural phase transitions and present rather complex crystalline
phases under compressions \cite{Neaton1999,Neaton2001,Hanfland2000}.
These structural changes are accompanied by a variety of intriguing
phenomena, among which the most striking is the anomalous melting
feature at high pressure. For sodium, its melting curve has been
measured up to 130 GPa \cite{Gregoryanz2005} and subsequently
reproduced by first-principles calculations based on either
molecular dynamics or the usual Lindemann criterion
\cite{Canales2008,Koci2008,Lepeshkin2009}. These studies have
revealed the unusual melting behavior of Na, i.e., the existence of
multiple maxima. For lithium, on the other hand, because of the
difficulties in containing the sample under high pressure, the
knowledge of its melting curve has long been confined to be less
than 8 GPa \cite{Luedemann1968,Boehler1983} until a recent
differential thermal analysis (DTA) measurement \cite{Lazicki2010},
which extended the melting line of Li up to 15 GPa and reported a
maximum at about 10 GPa. Theoretically, a Lindermann model curve of
Li was calculated \cite{Lepeshkin2009} to give an obvious
discontinuity near the bcc-fcc-liquid triple point, which was not
supported by experimental data \cite{Lazicki2010}. More directly,
Tamblyn \textit{et al.} \cite{Tamblyn2008} and
Hern$\acute{\texttt{a}}$dez \textit{et al.} \cite{Hernandez2010}
have from first-principles molecular dynamics (FPMD) simulations
determined the melting temperature of Li over a broad pressure
range. They both observed the negative slope of the melting curve.
In addition, a new phase in liquid lithium with $sp^{3}$ bonded
tetrahedral local order at pressures above 150 GPa was also
predicted \cite{Tamblyn2008}. A possible link between the maximum in
the melting line and liquid-liquid phase transition (LLPT) was
suggested for some systems, such as P \cite{Katayama2000,Monaco2003}
and Cs \cite{Falconi2005}. As for Na, first-principles simulations
have revealed that the maximum in the melting line occurs without
any accompanying LLPT, and higher compressibility of the liquid
phase than the solid phase causes the change of melting line slope
from positive to negative \cite{Eduardo2007}. The cause of the
anomalous melting behavior of Li still remains unclear. Though
previous FPMD studies have found bcc-like to fcc-like structural
transition in liquid Li \cite{Tamblyn2008,Yang2010}, the
temperatures are well above the melting temperature and thus it
could not be concluded that the similar structural transition occurs
along the melting line. It is desirable to explore whether LLPT
exists along the melting line and reveal what contributes to the
melting curve maximum of Li.

There are two common strategies for FPMD calculation of the melting
curve, i.e., the two-phase method \cite{Alfe2003,Belonoshko2005}
corresponding to a heterogeneous melting and the one-phase approach
\cite{Ogitsu2003} which involves a homogeneous melting. The first
one can give accurate results at the cost of computational
demanding, while the second usually causes superheating effect,
though needs much less atoms. A useful alternative has been proposed
by Belonoshko \textit{et al.} \cite{Belonoshko2006}, which is
referred to as \textit{Z}-method. It combines the advantages of both
techniques. When the system reaches the critical superheating
$T_{ls}$, it could not be heated before transforming into a liquid
structure. If let it evolve naturally as in \textit{Z}-method, the
temperature just drops down to the melting temperature $T_{m}$.
Along this line, recently, growing interests are concentrated on the
mechanism of homogeneous melting through characterizing the crystal
properties at $T_{ls}$
\cite{Belonoshko2007,Davis2009,Leighanne2010}. However, no consensus
could be reached. The dynamic and optical properties of crystal at
$T_{ls}$, which are deemed useful for insight into the mechanism of
melting, are still scarcely presented in the literature. Besides,
the optical properties should be different for the superheating
solid and the disordered liquid phase, and thus they could be able
to diagnose the homogeneous melting. The similar idea has been
suggested for diagnosing the shock melting of Al \cite{Ogitsu2009}.

Inspired by the above-mentioned facts, in this paper we calculate
the melting curve of Li up to 30 GPa using the \textit{Z}-method
implemented by FPMD simulations. We show that the shape of
\textit{Z} curve inverse with pressure increased to $\mathtt{\sim
}$8.2 GPa, which predicts the presence of maximum in the melting
curve, in good agreement with experimental measurement. Through
examining the static properties of Li at $T_{ls}$ and $T_{m}$, we
conclude that no LLPT occurs and liquid phase is more compressible
than solid phase, which may be the cause to the melting line
maximum. Besides, the dynamical and optical properties at $T_{m}$
and $T_{Ls}$ are studied. It is found that solid and liquid spectra
show marked difference. This could be an efficient way of diagnosing
the phase transition during the homogeneous melting. In the next
section, the methods used in homogeneous melting simulation and
optical properties calculations are described. In Sec. III, the
calculated results are discussed and compared with experimental
data. Finally, we close our paper with a summary of our main
results.

\section{Method}

We have performed FPMD simulations to determine the melting curve
using \textit{Z}-method, which has been proven successful to predict
melting temperature in several systems, such as Al
\cite{Bouchet2009}, H \cite{Davis2008}, MgO \cite{Belonoshko2010}
and so on. The idea is to perform FPMD in the microscopic ensemble
(NVE) on a single solid system at different initial $K$ (kinetic
energy). A realistic $T_{ls}$ can be reached without any external
intervention on the dynamics of the melting process. On further
increasing $K$ slightly, $T_{ls}$ will drop naturally to the melting
temperature $T_{m}$ at the pressure fixed by the chosen density. By
performing long microscopic simulations at different cell volumes,
it is possible to obtain points $\left(P,\, T_{m}\right)$ directly
on the melting curve. The method is as straightforward as the
two-phase method, and it requires half as many atoms in the
simulation cell. However, it still requires a large amount of
simulation steps to achieve complete melting curve.

We performed \textit{Z}-method simulations of Li melting with
\textit{ab initio} Simulation Package (VASP) \cite{Kresse1993} for
bcc structure (for eight densities). The all-electron projector
augmented wave (PAW) method \cite{Kresse1999,Bloch1994} was adopted,
retaining only the 2s electron in the valence, and the
exchange-correlation energy was described employing the generalized
gradient approximation (GGA) of Perdew-Burke-Ernzerhof (PBE)
formalism \cite{Perdew1996}, as implemented in VASP. We used
plane-wave cutoff of 150 eV, and the Brillouin zone was sampled only
with the $\Gamma$ point. For each density the system was simulated
for 10000-20000 steps with the time step of 2.0 fs, where the time
scale lies between 20 to 40 ps for different initial $K$ in order to
construct an isochoric curve $P$ versus $T$. In this study, although
the applied pressure range is up to 30 GPa, only the bcc structure
was used in the MD calculations. A 256-atom bcc supercell was
constructed.

Following the FPMD simulations, a total of ten configurations are
selected from an equilibrated (in an average sense) portion of the
molecular dynamics run, typically sampling the final picosecond of
evolution. For each of these configurations, the electrical
conductivity is calculated using the Kubo-Greenwood formula
\cite{Harrison1970}. The Kubo-Greenwood formulation gives the real
part of the electrical conductivity as a function of frequency
$\omega$,

\begin{eqnarray} \label{real-conductivity}
\sigma(\omega)&=\frac{2\pi}{3\omega\Omega}\sum\limits_{\textbf{k}}w(\textbf{k})\sum\limits_{j=1}^{N}\sum\limits_{i=1}^{N}\sum\limits_{\alpha=1}^{3}[f(\epsilon_{i,\textbf{k}})-f(\epsilon_{j,\textbf{k}})]
\cr
&\times|\langle\Psi_{j,\textbf{k}}|\nabla_{\alpha}|\Psi_{i,\textbf{k}}\rangle|^{2}\delta(\epsilon_{j,\textbf{k}}-\epsilon_{i,\textbf{k}}-\hbar\omega),
\end{eqnarray}
where $f(\epsilon_{i,\textbf{k}})$ describes the occupation of the
$i\textnormal{th}$ band, with the corresponding energy
$\epsilon_{i,\textbf{k}}$ and the wave function
$\Psi_{i,\textbf{k}}$ at \textbf{k}, and $w\left(\textbf{k}\right)$
is the \textbf{k}-point weighting factor.

\section{Results and Discussion}

\subsection{Melting curve}

By performing the microscopic \textit{Z}-method simulations, the
system evolve freely without any temperature control, and in each
case, after reaching $T_{ls}$, the isochore line drops to a point
$(P,\, T_{m})$ that should fall precisely along the melting curve.
The isochore plots for each density are shown in Fig. 1, from which
the melting points (the ones marked with square on each plot) and
critical superheating points are extracted and shown in Table I with
their respective error estimations.

\begin{figure*}
\includegraphics[clip,scale=0.3]{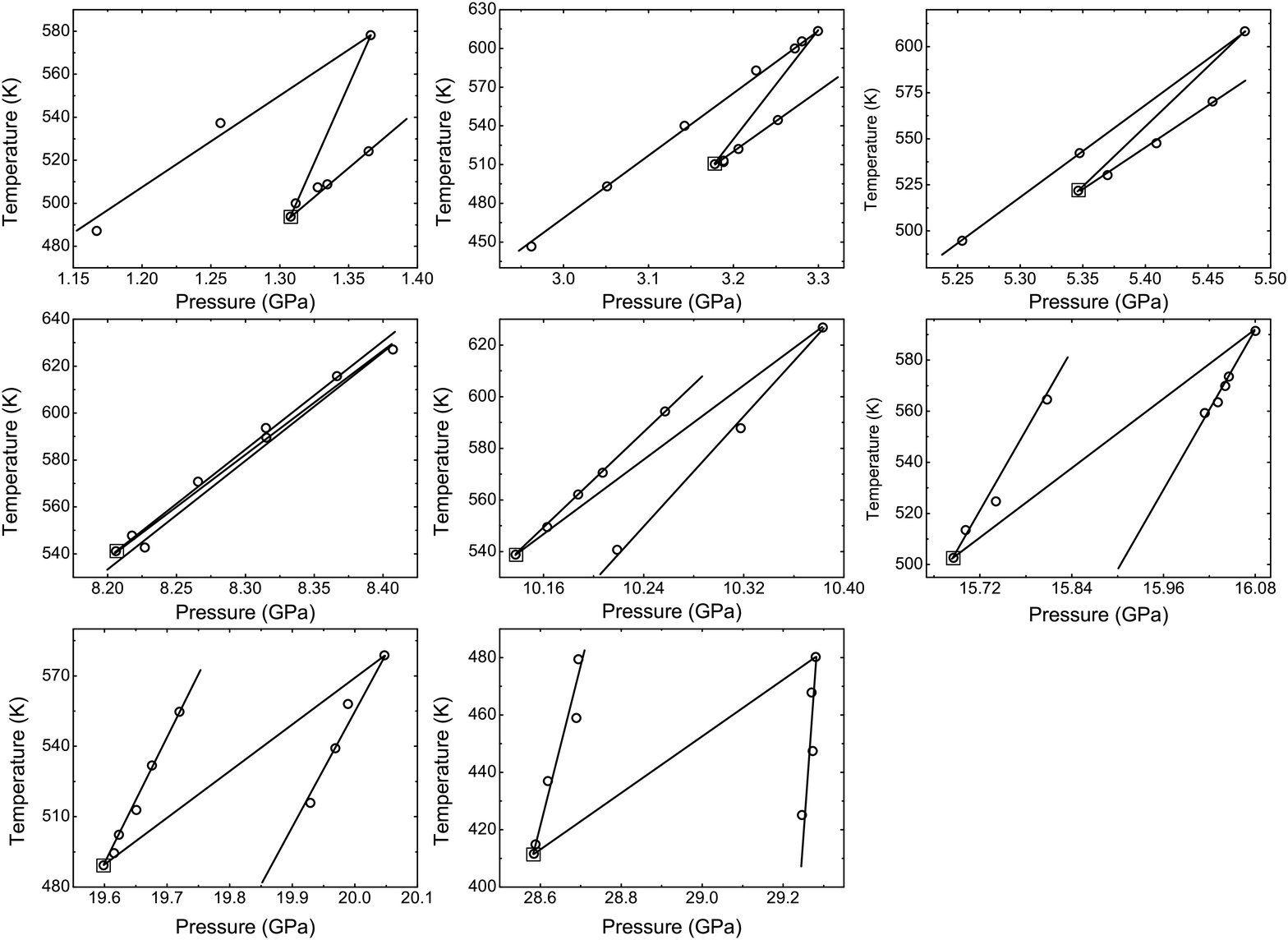}

\caption{Isochoric curves for $\rho=$ 0.58, 0.65, 0.71, 0.78, 0.82,
0.93, 1.0, and 1.13 g/cm$^{3}$. The respective melting points are
marked as squares in the different plots. Solid lines are just a
guide for eye.}

\end{figure*}

\begin{table}
\caption{Melting temperature $T_{m}$ and critical superheating
temperature $T_{ls}$ for Li at different pressures.}

\begin{longtable}{>{\centering}m{2.5cm}>{\centering}m{2.5cm}>{\centering}m{2.5cm}}
\hline \hline Pressure  & Melting temperature & Superheating
temperature\tabularnewline
\endhead
28.58$\pm$0.04 & 411.56$\pm$7.92 & 480.20$\pm$8.93\tabularnewline
\hline
\hline
\endlastfoot
(GPa) & (K) & (K)\tabularnewline
\hline
1.31$\pm$0.03 & 495.36$\pm$9.27 & 578.16$\pm$13.15\tabularnewline
3.18$\pm$0.03 & 510.29$\pm$9.73 & 613.38$\pm$12.46\tabularnewline
5.35$\pm$0.02 & 521.76$\pm$10.13 & 608.36$\pm$12.11\tabularnewline
8.21$\pm$0.02 & 541.15$\pm$10.68 & 627.1$\pm$13.16\tabularnewline
10.14$\pm$0.02 & 538.78$\pm$10.34 & 626.82$\pm$12.86\tabularnewline
15.69$\pm$0.02 & 502.62$\pm$9.21 & 591.42$\pm$12.02\tabularnewline
19.60$\pm$0.04 & 489.32$\pm$7.92 & 578.78$\pm$10.99\tabularnewline
\end{longtable}
\end{table}

As can be seen in the plots in Fig. 1, these eight isochores
naturally fall into two categories according to their characteristic
shapes. The first three form a {}``Z'' shape, while the last five
form an inverse {}``Z'' shape. The fourth plot should be noticed
that the upper and lower cap of {}``Z'' are nearly overlapped. These
featured isochore characteristics just predict the anomalous melting
behavior of lithium. When the volume is fixed at values
corresponding to the isochores with {}``Z'' shape, the pressure of
the liquid phase is higher than that of the solid phase at the
melting temperature. This implies that in condition of constant
pressure, the liquid phase would have a larger volume, and thus the
pressure derivative of the melting line in this region should be
positive according to the Clausius-Clapeyron equation

\begin{eqnarray}
\frac{dT_{m}}{dP} & = & T_{m}\frac{\vartriangle V}{\vartriangle
H},\end{eqnarray} where $T_{m}$ is the melting temperature, $P$ is
the pressure, $\vartriangle V=V_{l}-V_{s}$ is the difference of
molar volumes, and $\vartriangle H=H_{l}-H_{s}$ is the difference of
molar enthalpies. On the other hand, in the region of the isochores
with inverse {}``Z'', the case is totally opposite. That is, the
solid phase would have a larger volume, consistent with the negative
slope of the melting curve. In the case that the {}``Z'' shape
almost merges into one line as shown in the fourth plot in Fig. 1,
the melting temperature would reach its maximum. We further show the
pressure difference between the liquid ($P_{l}$) and solid ($P_{s}$)
phases as a function of volume at melting point in Fig. 2, which are
obtained from the isochore simulations. It is indicated that the
melting curve maximum locates at about 8.2 GPa, where $P_{l}$ equals
to $P_{s}$. This agrees well with the recent experimental
measurement \cite{Lazicki2010} of a negative slope above
$\mathtt{\sim }$1o GPa.

\begin{figure}
\includegraphics[clip,scale=0.4]{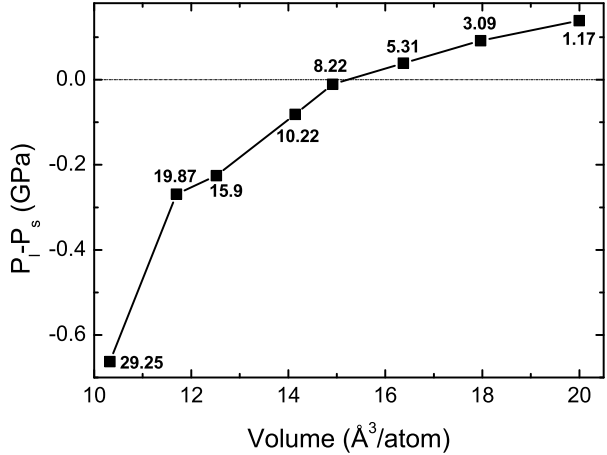}

\caption{Pressure difference between the liquid ($P_{l}$) and solid ($P_{s}$)
phases as a function of volume. The corresponding pressures are indicated.}

\end{figure}

The melting curve along with the estimated errors is shown in Fig.
3, in comparison with experimental data and data from two-phase FPMD
simulations. The maximum melting temperature of $\mathtt{\sim }$540
K is determined to be in the pressure range from 8 to 10 GPa. There
is reasonable agreement between the Z-method melting and
experimental results from Boehler \textit{et al.} \cite{Boehler1983}
and Luedemann \textit{et al.} \cite{Luedemann1968}, where the
discrepancy is below 20 K. Especially, they reports a melting
temperature of 508 K at 3 GPa, while we obtain 510 K at 3.18 GPa.
And also unlike other single-phase simulations of Li, the present
Z-method produces melting line very close to the two-phase
simulation results up to 30 GPa we considered. The two-phase
simulations have shown an interesting feature that bcc is more
stable than fcc structure close to melting \cite{Hernandez2010}.
Therefore, it should be reasonable to study the properties of
lithium at melting and superheating temperatures just based on the
simulation results of bcc structure in the following.

\begin{figure}
\includegraphics[clip,scale=0.4]{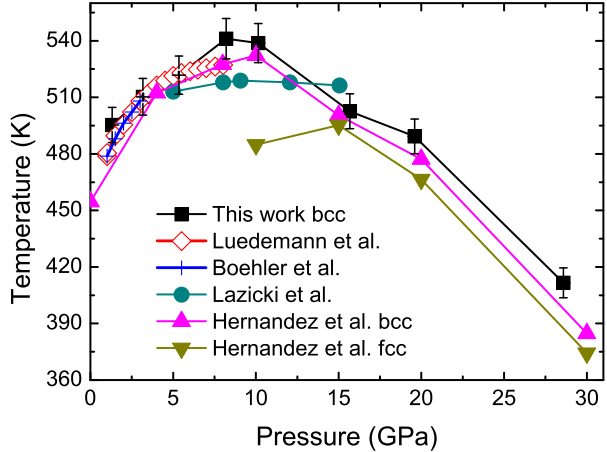}

\caption{(Color online) Melting curve for lithium up to 30 GPa. The
squares represent the individual points obtained with the
\textit{Z}-method. The diamonds are some experimental points from
Luedemann \textit{et al.} (Ref. 11) up to 8 GPa, crosses are
experimental points from Bochler \textit{et al.} (Ref. 12) up to 3
GPa, and circles are recent experimental points from Lazicki
\textit{et al.} (Ref. 13) up to 15 GPa. The up-triangles and
down-triangles are the theoretical points of bcc and fcc with
two-phase simulations, respectively (Ref. 15).}

\end{figure}

First-principles simulations have confirmed that the anomalous
melting behavior of Na is attributed to the high compressibility of
liquid phase than solid phase, and the maximum of the melting line
occurs without any accompanying first-order LLPT. Through analyzing
the pair correlation function (PCF) $g\left(r\right)$, we find that
these conclusions also hold for Li. As we all know, PCF is usually
used to examine atomic configurations, defined as $\rho
g\left(r\right)=\frac{1}{N}\left\langle \underset{i,j\neq
i}{\sum}\delta\left(r+\mathbf{r}_{i}-\mathbf{r}_{j}\right)\right\rangle
$, with $r$ the interatomic distance, $N$ the number of atoms,
$\rho$ the density $N/V$, and $\mathbf{r}_{i}$ and $\mathbf{r}_{j}$
the positions of atoms $i$ and $j$, respectively. In Fig. 4(a)
density dependence of PCF at $T_{m}$ and $T_{ls}$ are presented with
$r$ scaled by $r_{0}$, where $r_{0}$=$(N/V)^{1/3}$. As seen in the
figure, the shape of the PCF does not change up to 1.13 g/cm$^{3}$
($\mathtt{\sim }$30 GPa), which indicates that the compression in
this pressure range is uniform with the local structue unchanged.
However, the structure differences between $T_{ls}$ and $T_{m}$ are
noticeable especially for large values of $r$. For example, the peak
structure at around $r/r_{0}=2.8$ at $T_{ls}$ is large blurred at
$T_{m}$. Furthermore, the coordination at $T_{m}$ and $T_{ls}$ for
different densities are given from integrating $g\left(r\right)$
within a sphere of radius $R$, as shown in Fig. 4(b). In particular,
by integrating $g\left(r\right)$ to the first minimum, the average
coordination numbers $C_{NN}$ can be obtained, as indicated in the
Fig. 4(b). The first minimum of $g\left(r\right)$ at $T_{m}$ is at
$r/r_{0}\mathtt{\sim }1.5$, which is indicated by the dotted line.
It is obvious that $C_{NN}$ is nearly constant at 14 for these
densities, similar to that of a bcc crystal. Thus our simulations
give that no LLPT occurs along the melting line of Li up 30 GPa. It
is consistent with the experimental results \cite{Lazicki2010} that
no signature of bcc-fcc transition along the melting line was
observed up to 15 GPa, though liquid structural transformation from
bcc to fcc in pure liquid phase above melting temperature is a
natural case \cite{Yang2010}. In addition, as can be seen from Fig.
4(b), beyond the first coordination shell, the coordination number
for the liquid phase increases compared to the solid phase, which
implies that the liquid phase is more compressible than the solid
phase, and thus may lead to the maximum of the melting line.

\begin{figure*}
\includegraphics[clip,scale=0.35]{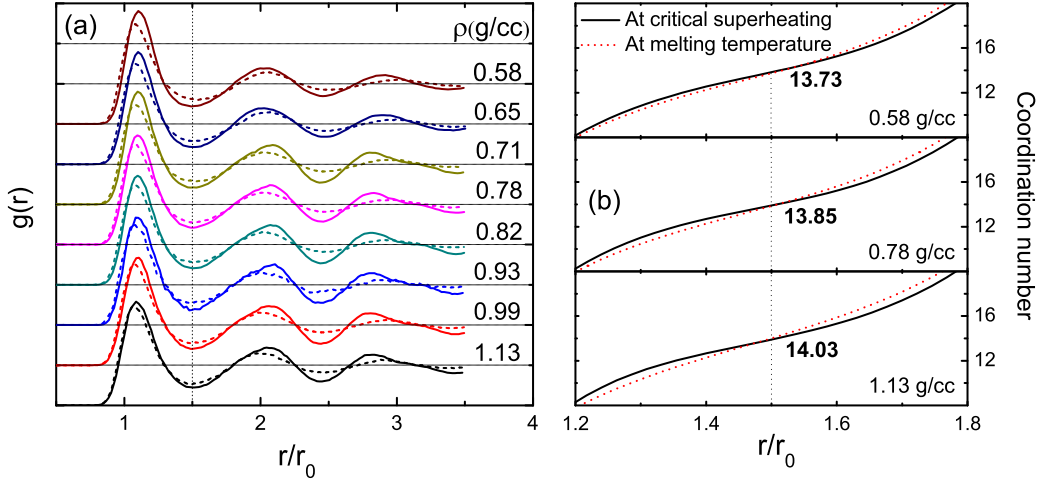}

\caption{(Color online) Density dependence of (a) pair correlation
function $g\left(r\right)$ and (b) coordination number at $T_{m}$
(solid line) and $T_{ls}$ (short dashed line). The vertical dotted
lines indicate the first minimum of $g\left(r\right)$. To facilitate
the comparison between different densities, $r$ is scaled by
$r_{0}$. }
\end{figure*}

Besides the above-presented clarification of the structure, we
further examine the mean square displacement (MSD) at $T_{m}$ and
$T_{ls}$ in order to confirm that we indeed have a solid behavior at
$T_{ls}$ and a liquid behavior at $T_{m}$ for each of the volumes
considered. Figure 5 shows the MSD up to 3 ps for $P\simeq10$ GPa at
$T_{ls}=627$ K and $T_{m}=539$ K. For purpose of comparison, we also
include the MSD at $T=540$ K (solid) and $T=571$ K (liquid). It is
obvious to see that the MSD at $T_{m}$ increases linearly at long
time, in a similar way as in pure liquid at $T=571$K, while at
$T_{ls}$ the displacement reaches a constant nearly, which suggests
that atoms are in a solid structure and could not diffuse away from
their equilibrium positions (compare with MSD at $T=540$ K).

\begin{figure}

\includegraphics[bb=0bp 0bp 567bp 404bp,clip,scale=0.38]{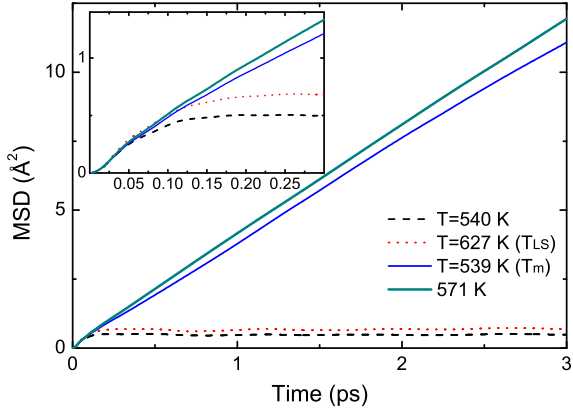}

\caption{(Color online) MSD of Li atoms for $P\simeq10$ GPa at
$T=540$ K (dashed line), $T_{ls}=627$ K (dotted line), $T_{m}=539$ K
(thin solid line), and $T=571$ K (thick solid line). Inset shows the
same curves up to 0.3 ps.}

\end{figure}

The diffusion coefficient $D$, estimated using the MSD up to
$t_{0}=0.5$ ps by

\begin{align}
D & =\frac{\left\langle r\left(t_{0}\right)^{2}\right\rangle
}{6t_{0}},\end{align} is shown in Fig. 6 for every pressure point in
our calculated melting curve. Here it is verified that there is a
remarkable difference in the atomic diffusion between at $T_{m}$ and
at $T_{ls}$ in the whole pressure range considered. As expected, the
diffusion decreases with pressure generally. Especially, the
diffusion coefficient at 1.31 GPa and $T_{m}=495$ K is about 0.69
$\lyxmathsym{\AA}^{2}/\textnormal{ps}$, which is in excellent
agreement with the experimental value of $0.69\pm0.09$
$\lyxmathsym{\AA}^{2}/\textnormal{ps}$ at 0 GPa and $470$ K
\cite{Lowenberg1967}.

\begin{figure}
\includegraphics[clip,scale=0.38]{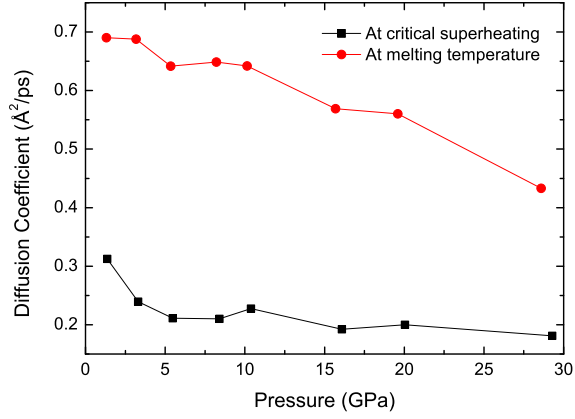}

\caption{(Color online) Diffusion coefficient for Li atoms compared
for each pressure at melting point (circles) and superheating point
(squares).}

\end{figure}

\subsection{Dynamic conductivity and optical properties}

The linear optical conductivities of superheated and melted Li at
different densities (0.58, 0.78 and 1.13 g/cm$^{3}$) calculated
using Eq. (1) are shown in Fig. 7, which are averaged over ten
snapshots selected during the course of FPMD simulations. For
identical density, the spectra at $T_{ls}$ and $T_{m}$ show marked
differences. At $T_{ls}$, there are some structural peaks, while the
dips fill in and only leave a shoulder at $T_{m}$. Of note is the
fact that our predicted difference between superheated solid and
melting liquid is even more pronounced at lower densities. For
example, at density of 0.58 and 0.78 g/cm$^{3}$, there are two
prominent peaks, while only one at 1.13 g/cm$^{3}$. In addition, for
critical superheated solid, the peaks broaden while moving to higher
energy as density is increased. All of the superheated solid samples
exhibit nearly free-electron characters. Again, the liquid spectrum
is featureless and Drude-type. This allows us to expect the optical
measurement to be able to diagnose the homogeneous melting.

\begin{figure}
\includegraphics[bb=0bp 0bp 566bp 484bp,clip,scale=0.4]{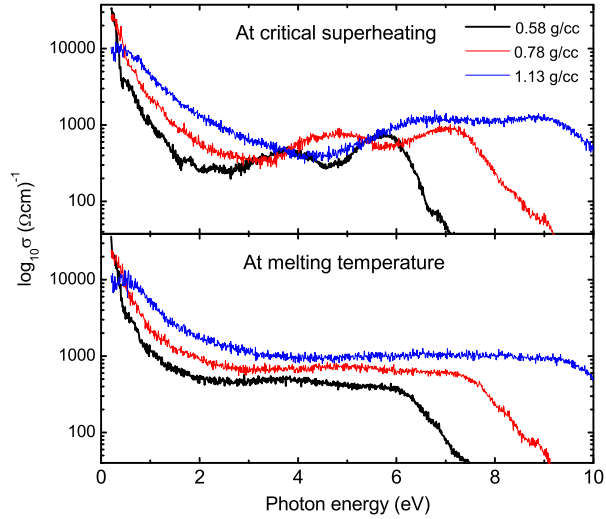}

\caption{(Color online) The lithium optical conductivity versus
$\omega$ at (a) critical superheating temperature and (b) melting
temperature for different densities of 0.58, 0.78 and 1.13
g/cm$^{3}$. Logarithm longitudinal scale is used.}

\end{figure}

Through extrapolating to the zero frequency limit, the dc
conductivity can be determined by fitting with the simple Drude form
\cite{Harrison1970}

\begin{eqnarray}
\sigma\left(\omega\right) & = &
\frac{\sigma_{\textnormal{dc}}}{1+\omega^{2}\tau_{D}^{2}},\end{eqnarray}
where $\tau_{D}$ represents effective collision time. Figure 8
presents the dc conductivity of lithium at $T_{ls}$ and $T_{m}$ for
different densities. The dc conductivity shows a systematic behavior
in terms of density for both cases. Stronger ion-electron scattering
with increasing density would diminish the conductivity. In
addition, the conductivity decreases with the temperature for
metals. It is thus natural that the conductivity decrease with
increasing density in the region with positive melting line slope.
However, in the region with negative melting line slope, the effect
of increased scattering is more prominent than that of the decreased
temperature, which plays a central role in diminishing the
conductivity. The dc conductivities at $T_{m}$ are in order of
$10^{4}\:\left(\Omega\textnormal{cm}\right)^{-1}$, which is
consistent with the experimental and other theoretical results of
the liquid lithium in the similar density and temperature ranges
\cite{Bestea2002,Kiezmann2008}. Also it can be seen from Fig. 8 that
the dc conductivity difference between $T_{m}$ and $T_{ls}$ becomes
smaller with increasing density, and even undistinguished at
1.13g/cm$^{3}$.

\begin{figure}
\includegraphics[clip,scale=0.4]{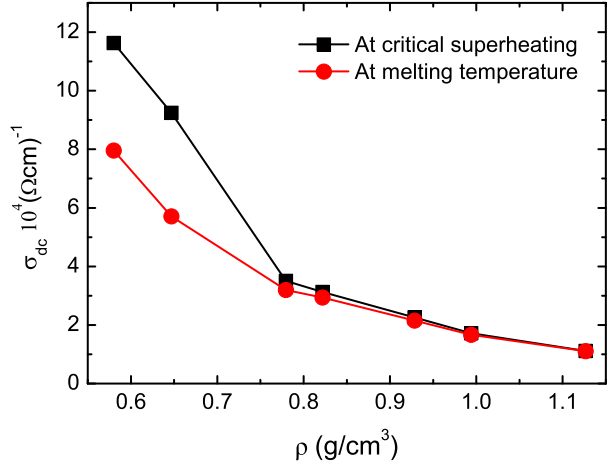}

\caption{(Color online) Electron dc conductivities at critical
superheating temperatures (squares) and melting temperatures
(circles) for different densities.}

\end{figure}

\section{Conclusion}

In summary, we have performed FPMD simulations of the melting curve
of lithium up to 30 GPa using \textit{Z}-method. The results are in
good agreement with experimental measurements and the two-phase
simulations. It can be concluded that the melting line maximum of
lithium may be caused by higher compressibility of liquid phase than
solid phase, without LLPT accompanied. In addition, we have also
systematically studied the atomic dynamic diffusion behavior and
electronic dynamic conductivity properties at the critical
superheating and melting points of lithium, which have revealed
prominent physical differences between the superheated solid phase
and the disordered liquid phase. For these two homogeneous phases,
interestingly, the electron conductivities, especially the dc
components, show the merging tendency at high densities, which
suggests the increasing role the local structure plays in
determining the electron-ion scattering.

\begin{acknowledgments}
One of the authors (D.L.) is grateful for helpful discussions with
H. Liu. This work was supported by NSFC under Grants No. 51071032
and No. 10734140, by the National Basic Security Research Program of
China, and by the Foundations for Development of Science and
Technology of China Academy of Engineering Physics under Grant No.
2009B0301037.
\end{acknowledgments}

\end{document}